\documentclass[twocolumn,english,aps,pra,showpacs,superscriptaddress]{revtex4-1}
\usepackage[T1]{fontenc}
\usepackage[utf8]{inputenc}
\usepackage{booktabs}
\usepackage{amsmath}
\usepackage{amssymb}
\usepackage{graphicx}

\makeatletter

\providecommand{\tabularnewline}{\\}

\@ifundefined{textcolor}{}
{%
 \definecolor{BLACK}{gray}{0}
 \definecolor{WHITE}{gray}{1}
 \definecolor{RED}{rgb}{1,0,0}
 \definecolor{GREEN}{rgb}{0,1,0}
 \definecolor{BLUE}{rgb}{0,0,1}
 \definecolor{CYAN}{cmyk}{1,0,0,0}
 \definecolor{MAGENTA}{cmyk}{0,1,0,0}
 \definecolor{YELLOW}{cmyk}{0,0,1,0}
}

\usepackage{amsfonts}

\def\be{\begin{equation}}
\def\ee{\end{equation}}
\def\bea{\begin{eqnarray}}          
\def\eea{\end{eqnarray}}
\def\bi{\begin{itemize}}
\def\ei{\end{itemize}}

\usepackage{babel}

\makeatother

\usepackage{babel}
\begin{document}

\title{Second-order Peierls transition\\
 in the spin-orbital Kumar-Heisenberg model }

\author{Wojciech Brzezicki}

\affiliation{Marian Smoluchowski Institute of Physics, Jagiellonian University,
             prof. S. \L{}ojasiewicza 11, PL-30348 Krak\'ow, Poland }

\affiliation{CNR-SPIN, IT-84084 Fisciano (SA), Italy, and \\
             Dipartimento di Fisica \textquotedbl{}E. R. Caianiello\textquotedbl{},
             Universit\'a di Salerno, IT-84084 Fisciano (SA), Italy}

\author{Imre Hagymási}

\affiliation{Strongly Correlated Systems \textquotedbl{}Lendület\textquotedbl{}
Research Group, Institute for Solid State Physics and Optics, MTA
Wigner Research Centre for Physics, Budapest H-1525 P.O. Box 49, Hungary}

\affiliation{Department of Theoretical Physics, University of Szeged, Tisza Lajos
krt. 84-86, H-6720 Szeged, Hungary}

\author{Jacek Dziarmaga}

\affiliation{Marian Smoluchowski Institute of Physics, Jagiellonian University,
             prof. S. \L{}ojasiewicza 11, PL-30348 Krak\'ow, Poland }

\author{Örs Legeza}

\affiliation{Strongly Correlated Systems \textquotedbl{}Lendület\textquotedbl{}
Research Group, Institute for Solid State Physics and Optics, MTA
Wigner Research Centre for Physics, Budapest H-1525 P.O. Box 49, Hungary}

\begin{abstract}
We add a Heisenberg interaction term $\propto\lambda$ in the one-dimensional SU(2)$\otimes$XY spin-orbital model introduced by B. Kumar.
At $\lambda=0$ the spin and orbital degrees of freedom can be separated by a unitary transformation leading to an exact solution of the model. 
We show that a finite $\lambda>0$ leads to spontaneous dimerization of the system 
which in the thermodynamic limit becomes a smooth phase transition at $\lambda\to 0$, 
whereas it remains discontinuous within the first order perturbation approach. 
We present the behavior of the entanglement entropy, energy gap and dimerization order parameter in the limit of $\lambda\to 0$ confirming the critical behavior. 
Finally, 
we show the evidence of another phase transition in the Heisenberg limit, $\lambda\to\infty$, 
and give a qualitative analytical explanation of the observed dimerized states both in the limit of small and large $\lambda$.
\end{abstract}


\maketitle

\section{Introduction}

A Peierls transition being a distortion of the periodic lattice of
a one-dimensional (1D) crystal has been known for more than $80$
years \cite{Pei55}. Peierls' theorem states that a 1D equally spaced
chain with one electron per ion is unstable to the dimerization, i.e.,
a more favorable configuration can be achieved if every ion moves
closer to its one neighbor and further away from the other. This is
a purely kinetic effect related with the nesting of the 1D Brillouin
zone. Such dimerization involves charge and lattice degrees of freedom
but a similar process can be observed in the spin-orbital models
\cite{PhysRevLett.101.157204,PhysRevB.82.035108} where the interaction
within the orbital sector dimerize the spin order. Such an effect
was observed by a neutron scattering in the perovskite vanadate ${\mathrm{Y}\mathrm{V}\mathrm{O}}_{3}$
\cite{PhysRevLett.91.257202} as theoretically described by Horsch
and coworkers \cite{Hor03}. Similarly, for an iridate triangular-lattice
superconductor Ir$_{1-x}$Pt$_{x}$Te$_{2}$ the photoemission and
model studies have revealed that an orbitally induced Peierls effect
governs the charge and orbital instability in this compound \cite{PhysRevB.86.014519}.
A strict, 1D spin-orbital analog of the Peierls' dimerized state was
found in ${\mathrm{NaTiSi}}_{2}{\mathrm{O}}_{6}$ \cite{PhysRevB.69.020409},
as shown by Raman scattering measurements, where the dimerized spin
state at low temperature is driven by the orbital fluctuations at
high temperature. A quite different situation emerges when the orbital
fluctuation do not drive the spins' dimerization but the orbital order
makes the electrons' hopping quasi-1D so that the classical Peierls
transition can occur \cite{PhysRevB.89.161112}. Such an orbital-selective
Peierls dimerization was found in the $t_{2g}$ spinel ${\mathrm{MgTi}}_{2}{\mathrm{O}}_{4}$
by an optical measurements. On the other hand it was shown that the
notion of the Peierls instability also applies to the strongly interacting
electrons coupled to the phonons within the Holstein model \cite{PhysRevB.75.035124}
which allows to explain the orbital order in the undoped manganite
${\mathrm{LaMnO}}_{3}$ \cite{PhysRevB.80.235123}.

Spin-orbital models are crucial from the point of view of frustrated
magnetism \cite{Nag00,Hfm,Kha05,*Ole05,*Ole12} as the orbital interactions,
being typically of lower symmetry than the SU(2)-symmetric spin interactions,
can lead to frustration in both sectors even on a square lattice.
This often happens in the transition-metal oxides. When the $3d$
orbitals are partly filled, electrons localize due to large on-site
Coulomb interaction and superexchange between magnetic ions includes
both spin and orbital degrees of freedom that are strongly interrelated
\cite{Kug82}. The orbital degeneracy leads then in many cases to
a dramatic increase of quantum fluctuations \cite{Fei97}, which may
trigger exotic order \cite{Brz12,Brz13}, or may stabilize a spin-liquid
\cite{PhysRevB.80.064413,*PhysRevB.87.224428,*PhysRevX.2.041013,Nor08,*Chal11,*Nor11}
when different states compete near a quantum critical point. In general
these fluctuations make the ground-states of the spin-orbital models
entangled, as it happens in the archetypal Kugel-Khomskii model in
one-, two-, three-dimensional and bilayer cases, respectively \cite{PhysRevB.86.224422,Brz12,Brz13,Brz11}.
Similarly, in the $S=1/2$ SU(2)$\otimes$SU(2) chain \cite{PhysRevB.61.6747,PhysRevB.58.10276},
both ground state \cite{Che07} and excited states \cite{You12} are
entangled and the $S=1$ SU(2)$\otimes$SU(2) chain which plays a
prominent role in the vanadium perovskites \cite{Kha05,*Ole05,*Ole12,Kha01,*Ole06,*Hor08}.
In exceptional cases can such 1D models be solved exactly, for example
at the SU(4) point \cite{PhysRevLett.81.3527,*PhysRevLett.82.835,*PhysRevLett.83.624}
or for a valence-bond state \cite{Ole07} of alternating spin and
orbital singlets similar to the Majumdar-Ghosh state in a 1D $J_{1}$-$J_{2}$
spin chain \cite{Maj69}, but even in these situations the spins and
orbitals cannot be separated from each other.

In real materials the symmetry between spin and orbital interactions
is absent. Orbital interactions generically have lower symmetry than
spin ones \cite{vdB99}, being usually Ising- or XY-like \cite{PhysRevB.61.5868}.
The XY case is quantum and in general the orbitals cannot be separated
from the spins \cite{PhysRevB.83.245130}. Therefore the 1D SU(2)$\otimes$XY
model introduced by Kumar \cite{PhysRevB.87.195105,*PhysRevB.77.205115,*PhysRevB.79.155121}
is very exceptional because; ($i$) it is exactly solvable and ($ii$)
by a change of basis, the $S=1/2$ spins decouple from the orbitals
in an \textit{open} chain. The orbital interactions remain formally
unchanged but the spin ones are gauged away. The spins then appear
free and the ground state has large degeneracy ($2^{L}$ for chain
length $L$) \cite{PhysRevB.87.195105}.

Quite surprisingly, the properties of the Kumar model are determined
by topology. When the chain is closed and the model becomes periodic
the exact solution still exists, as shown in Ref. \cite{Brz14prl}, 
but the spins can no longer be gauged away completely.
Contrarily, the gauge is accumulated on the closing orbital bond so
that the whole spin sector enters the orbital problem only through
the boundary condition, in analogy to Aharonov-Bohm magnetic flux
through the ring. This partially lifts the macroscopic degeneracy
of the ground state of the open chain and splits the manifold of the
$2^{L}$ states into a multiplet of low-lying states. These are
the states with topological excitations of a non-local nature having
quadratic dispersion in the total quasimomentum carried by the spins
\cite{Brz14prl}. Such a multiplet of states connected by the non-local
excitations is a characteristic feature of another model originating
from the orbital exchange, namely the two-dimensional orbital compass
model (OCM) \cite{Mil05,*Brz10,*Brz13b}, which was considered, iter
alia, in the context of topologically protected qubits \cite{Dou05}.
From this point of view it was important to study how the small admixture
of the SU(2) Heisenberg interactions, which is unavoidable in the
physical implementations employing arrays of Josephson junctions,
will modify the multiplet structure in OCM \cite{Tro10,*Tro12}.

In this paper we will study the spin-orbital Kumar model with an extra
antiferromagnetic (AF) Heisenberg term for the nearest-neighbor (NN)
spins, namely the Kumar-Heisenberg (KH) model. As we will see further
on, this model is no longer exactly solvable but the Kumar gauge transformation
can be still applied. As a result the pure Kumar term gets completely
or partially decoupled depending on the topology of the system but
the Heisenberg term is subject to the spin permutation that depends
on the configuration of orbitals in their canonical basis. In other
words, under the transformation the NN spin interaction gets smeared
out and this smearing is governed by the delocalization of the up/down
orbitals caused by the quantumness of the XY orbital interactions.
This situation is interesting because in this way the spin interactions
can get frustrated - the smeared interaction remains AF but now the
further neighbors are involved. On the other hand, as the pure Kumar
model leaves the spins practically completely uncorrelated, we expect
that any small admixture of spin interactions $\lambda$ will order
the spins in some way. This is a favorable situation as it will enable
us to use perturbation expansion in the manifold of degenerate ground-states
of pure Kumar model. 
To examine the region, where the coupling
between the spins cannot be treated as a small perturbation, we use
the density matrix renormalization group method (DMRG) \cite{white1992,schollwock2005,manmana2005,hallberg2006,schollwock2011},
which gives an accurate description of the ground state in one dimension.
This also provides an independent way to verify the perturbational
results. 

The paper is organized as follows. In Sec. \ref{sec:KHmodel} we introduce
the Kumar-Heisenberg (KH) model, in Sec. \ref{sec:Solution} we show
the exact solution of the periodic KH model based on Ref. \cite{Brz14prl},
in Sec. \ref{sec:ferro} we apply this solution to the ferromagnetic, $\lambda<0$,
case and get the exact ground state, and in Sec. \ref{sec:pert} we explain the
spin dimerization within a first order expansion in a small $\lambda>0$ which is a
valid description in the limit of short chains. In Sec. \ref{sec:crossover}
we explain why this limit cannot describe an infinite system where a second-order
phase transition occurs, as shown in Sec. \ref{sec:weaklambda}. In Sec. 
\ref{sec:Dimerization} we describe a dimerized state and give a 
mean-field mechanism of the orbital dimerization for large $\lambda$. In this
Section we also show the evidence of a second phase transition in the limit of 
$\lambda\to\infty$. The conclusions are presented in Sec. \ref{sec:Summa}
and in Appendix \ref{app:J} we give further details on the 
effective spin couplings introduced in Sec. \ref{sec:pert}.

\section{Kumar-Heisenberg model\label{sec:KHmodel}}

The Hamiltonian of the KH model reads 
\begin{eqnarray}
{\cal H} & = & {\cal H}_{0}+{\cal V},
\end{eqnarray}
where 
\begin{equation}
{\cal H}_{0}=J\sum_{l=1}^{L}X_{l,l+1}\!\left(\tau_{l}^{+}\tau_{l+1}^{-}\!+\!\tau_{l}^{-}\tau_{l+1}^{+}\right)\label{eq:ham_pbc}
\end{equation}
is the Kumar model and 
\begin{equation}
{\cal V}=\lambda\sum_{l=1}^{L}\!\vec{\sigma}_{l}\vec{\sigma}_{l+1}\label{V}
\end{equation}
is the Heisenberg term. Here $\tau$'s are orbital Pauli matrices
and $X_{i,j}\!=\!\left(1\!+\!\vec{\sigma}_{i}\vec{\sigma}_{j}\right)/2$
is a \textit{spin transposition} operator, defined by another set
of Pauli matrices $\sigma$. In the case of periodic boundary conditions
(PBC) $\vec{\sigma}_{L+1}=\vec{\sigma}_{1}$ and $\tau_{L+1}=\tau_{1}$.
For open boundary conditions (OBC) $\vec{\sigma}_{L+1}=0$ and $\tau_{L+1}=0$.

The Hamiltonian can be transformed into base where ${\cal H}_{0}$
appears as a purely orbital contribution. For this we use the unitary
operator 
\begin{equation}
{\cal U}=\prod_{l=1}^{L-1}\left[\frac{1-\tau_{l+1}^{z}}{2}+\frac{1+\tau_{l+1}^{z}}{2}\chi_{l+1,l}\right],\label{eq:U}
\end{equation}
where $\chi_{l+1,l}$ is a {\it spin permutation} operator composed of 
the spin transpositions $X_{i,j}$: 
\begin{equation}
\chi_{l+1,l}= X_{l+1,l}X_{l,l-1}...X_{3,2}X_{2,1}.
\end{equation}
Now according to Ref. \cite{Brz14prl} the transformed Hamiltonian
takes a form 
\begin{equation}
{\cal H}'\!=\!{\cal U}^{\dagger}{\cal H}\:{\cal U}\!=\!
{\cal H}'_{0}+{\cal V}',\label{eq:ham_tr}
\end{equation}
where 
\begin{eqnarray}
{\cal H}'_{0}\!=\! J\!\sum_{l=1}^{L-1}\!\left(\tau_{l}^{+}\tau_{l+1}^{-}\!+\!{\rm h.c.}\right)\!+\! 
J\!\left(\!R_{1}^{(1)}R_{1}^{(2)}\tau_{1}^{+}\tau_{L}^{-}\!+\!{\rm h.c.}\!\right)\!,\label{eq:H0'}
\end{eqnarray}
and 
\begin{equation}
{\cal V}'\!=\!\lambda\!\sum_{l=1}^{L}\!\vec{\sigma}_{l}\vec{\sigma}_{l+1}.\label{eq:V'}
\end{equation}
Here $R_{1}^{(1)}$ is a cyclic permutation of spins at sites $l=1,\dots,N$
by $1$ site and $R_{1}^{(2)}$ is the same permutation of spins at
sites $l=N+1,...,L$: 
\begin{eqnarray}
 &  & R_{1}^{(1)}:\{\vec{\sigma}_{1},...,\vec{\sigma}_{N}\}\to\{\vec{\sigma}_{2},...,\vec{\sigma}_{N},\vec{\sigma}_{1}\},\nonumber \\
 &  & R_{1}^{(2)}:\{\vec{\sigma}_{N+1},...,\vec{\sigma}_{L}\}\to\{\vec{\sigma}_{N+2},...,\vec{\sigma}_{L},\vec{\sigma}_{N+1}\}.
\end{eqnarray}
Here $N$ is a good quantum number of ''up'' ($\tau_{l}^{z}=+1$)
orbitals: 
\begin{equation}
N=\sum_{l=1}^{L}\frac{1+\tau_{l}^{z}}{2}.
\end{equation}
The primed spins $\vec{\sigma}'_{l}$ are entangled spin-orbital operators:
\begin{eqnarray}
\vec{\sigma}'_{l} & = & \left|\vec{\tau}^{z}\right\rangle \vec{\sigma}_{f(l,\vec{\tau}^{z})}\left\langle \vec{\tau}^{z}\right|,\label{eq:Uspin}
\end{eqnarray}
with $\left|\vec{\tau}^{z}\right\rangle =\left|\tau_{1}^{z},\tau_{2}^{z},\dots,\tau_{L}^{z}\right\rangle $
and a permutation of chain sites 
\begin{equation}
f(l,\vec{\tau}^{z})=l+\frac{1}{2}\left(1+\tau_{l}^{z}\right)\left(1-l\right)+\sum_{p=l+1}^{L}\frac{1}{2}\left(1+\tau_{p}^{z}\right).\label{f}
\end{equation}
$\vec{\sigma}'_{l}$ are the original spins $\vec{\sigma}_{l}$ after
a permutation of the chain sites dictated by the orbital configuration
in the $\tau_{l}^{z}$-eigenbasis.

\section{Solution of the Kumar model\label{sec:Solution}}

The Hamiltonian ${\cal H}'_{0}$ is exactly solvable \cite{Brz14prl}.
On all bonds but the closing one it involves only orbital degrees
of freedom and even on the closing bond the spin part affects only
the boundary conditions via the cyclic permutations $R_{1}^{(1,2)}$.
These can be diagonalized and turned to the phase factors $e^{i{\cal K}_{1,2}}$
with quasi-momenta (spin currents) ${\cal K}_{1}=\frac{2\pi n_{1}}{N}$
and ${\cal K}_{2}=\frac{2\pi n_{2}}{L-N}$. Here $n_{1}=0,\dots,N-1$
and $n_{2}=0,...,L-N-1$. Then the diagonalization of ${\cal H}'_{0}$
can be completed by the Jordan-Wigner (JW) transformation of the form
$\tau_{l}^{z}=1-2n_{l}$ and $\tau_{l}^{+}=c_{l}\prod_{j<l}(1-2n_{j})$,
where $n_{l}=c_{l}^{\dagger}c_{l}$ and $c_{l}$ annihilates a JW
fermion. The last step is a Fourier transformation, we have 
\begin{equation}
{\cal H}'_{0}=J\sum_{l=1}^{L-1}c_{l+1}^{\dagger}c_{l}+Je^{-2\pi i\Phi}c_{1}^{\dagger}c_{L}+{\rm h.c.}=2J\sum_{k}c_{k}^{\dagger}c_{k}\cos k,\label{eq:Hzero}
\end{equation}
where the phase $2\pi\Phi={\cal K}_{1}+{\cal K}_{2}-\pi(L-N-1)$ is
twisting the boundary condition, $c_{L+1}=e^{2\pi i\Phi}c_{1}$, just
like a magnetic flux $\Phi$ through the periodic ring. The standard
cosine dispersion of the fermions involves quasi-momenta $k$ quantized
as $k=\frac{2\pi}{L}\left(m+\Phi\right)$ with $m=0,...,L-1$.

As shown in Ref. \cite{Brz14prl}, for PBC the ground state (GS) of
the model is the Fermi sea of the JW fermions at half-filling, 
\begin{equation}
N=L/2,
\end{equation}
with a zero spin current, 
\begin{equation}
{\cal K}_{1}+{\cal K}_{2}=0.
\end{equation}
Thus the GS wave function of ${\cal H}'_{0}$ factorizes into a product
of an orbital and spin state: 
\begin{equation}
\left|0\right\rangle =c_{k_{1}}^{\dagger}\dots c_{k_{N}}^{\dagger}\left|vac\right\rangle _{c}\otimes\left|{\cal K}_{1}\right\rangle ^{(1)}\otimes\left|-{\cal K}_{1}\right\rangle ^{(2)},
\end{equation}
with $\left|vac\right\rangle _{c}$ being a vacuum of the JW fermions.
The orbital state is a Fermi sea with occupied quasimomenta $k_{j}=\frac{\pi}{N}(-j+\frac{1}{2})-\frac{\pi}{2}$
for $j=1,\dots,N$ such that $\cos k_{j}<0$. The spin states $\left|{\cal K}\right\rangle ^{(1,2)}$
are the ${\cal K}$-momentum eigenstates of the operators $R_{1}^{(1,2)}$
with an eigenvalue $e^{i{\cal K}}$. The ground state is degenerate
with respect to the $N$ values of ${\cal K}_{1}$.

For OBC the spin state is completely arbitrary and the orbital state
is a half-filled Fermi sea with OBC. Since in this paper we are interested
in the thermodynamic limit, from now on we focus on OBC.

The Kumar model is the special case of the Heisenberg coupling $\lambda=0$
in Eq. (\ref{V}). Its $2^{L}$-fold degenerate ground state is very
sensitive to the perturbation. In the following we scan different
regimes of $\lambda$.

\section{Ferromagnetic coupling: ${\bf \lambda<0}$ \label{sec:ferro}}

This case is easy to solve in the Kumar basis, where the ferromagnetic
spin state is the ground state of both the Heisenberg term (\ref{eq:V'})
and the Kumar model (\ref{eq:H0'}). The orbital ground state is a
half-filled Fermi sea.

Back in the physical representation, permuted spins are still ferromagnetic,
the exchange operators $X_{i,j}=1$, and the orbitals are in the ground
state of the anti-ferromagnetic XX-chain, see Eq. (\ref{eq:ham_pbc}).
The ground state is a product of the ferromagnetic spin state (FM) and
the orbital state - there is no spin-orbital entanglement.

\section{Perturbatively weak coupling: ${\bf 0<\lambda\ll1/L}$\label{sec:pert}}

On the antiferromagnetic side the solution is no longer so simple.
In order to get a rough idea how the highly degenerate Kumar ground
state may respond to a weak Heisenberg perturbation, it is tempting
to try a perturbation theory in weak $\lambda$. In the Kumar representation
for OBC, the spin state is arbitrary and the orbital state is a half-filled
Fermi sea with a gap $\simeq1/L$. When the perturbation is weaker
than the orbital gap, $0<\lambda\ll1/L$, one can treat the spins
with the degenerate perturbation theory.

To first order in $\lambda$, the orbital state does not change, protected
by the orbital gap, and the spin degeneracy is removed by a Hamiltonian
\begin{equation}
{\cal H}'_{1}=\langle{\cal V}'\rangle=\lambda\sum_{i,j=1}^{L}\frac{1}{2}\mbox{J}_{i,j}~\vec{\sigma}_{i}\vec{\sigma}_{j},\label{eq:H1J-obc}
\end{equation}
where the average $\langle..\rangle$ is taken in the orbital Fermi
sea. In the physical representation only nearest-neighbor (NN) spins
are coupled.

\begin{figure}[t!]
\includegraphics[clip,width=1\columnwidth]{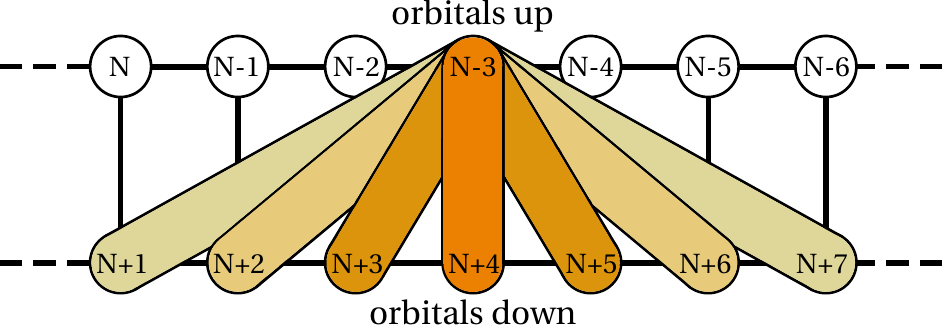}
\caption{\label{fig:ladder} The schematic view of the bonds connecting the
two spin subsystems (upper and bottom legs of the ladder) in the first-order
Hamiltonian in Eq. (\ref{eq:H1J-obc}). The spin bonds connecting
site $l=4$ of the orbital-up subsystem with sites of the orbital-down
subsystem are marked. Their color intensity increases with the strength
of the coupling $\mbox{J}_{i,j}$. }
\end{figure}

After the Kumar transformation, each physical NN-coupling contributes
to (is smeared over) many $\mbox{J}_{i,j}$ (see Appendix \ref{app:J}
for more details). The transformation is a permutation that maps the
$N$ consecutive spins on empty sites (orbitals up) to spins $i=N,...,1$
and the $N$ consecutive spins on occupied sites (orbitals down) to
spins $i=N+1,...,2N$. The exact matrix $\mbox{J}_{i,j}$ is described
in more detail in Appendix \ref{app:J}. It turns out to be dominated
by the couplings between the two $N$-spin subsystems, as shown in
Fig. \ref{fig:ladder}.
\begin{figure}[t!]
\includegraphics[clip,width=1\columnwidth]{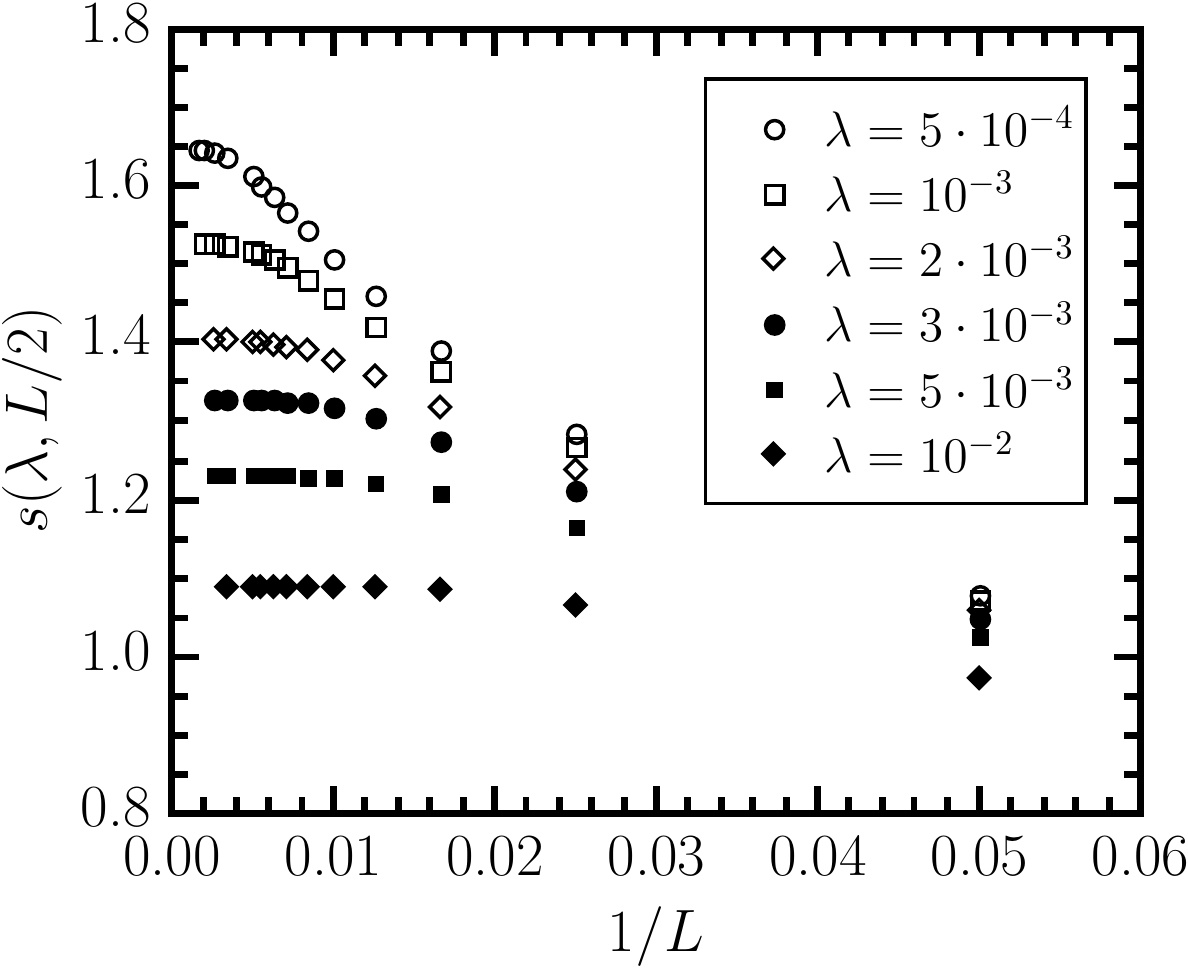}
\caption{\label{fig:block_entropy_scaling} Finite-size scaling results for
the block entropy of an open chain as a function of $1/L$ (up to length $L=600$) for several
values of $\lambda$. For each $\lambda$ the entropy saturates for
small enough $1/L$ where the thermodynamic limit is achieved. }
\end{figure}
The strongest bonds are along the rungs of
the spin ladder. When we keep only them, then the ground state will
become a product of spin singlets along the rungs. We compared the
energy of this trial state to the energy from ${\cal H}'_{1}$ with
the full matrix $\mathbf{J}$. For $L=16$ the energies differ by
less than $0.00004$ in case of $\lambda=0.01$ and by slightly more
than $0.0003$ for $\lambda=0.1$.

A full trial state is the spin trial state times the orbital Fermi
sea. It can be transformed back to the physical representation. For
$L=16$ the result shows strong dimerization of spins i.e., $\langle\vec{\sigma}_{l}\vec{\sigma}_{l+1}\rangle<0$
on odd NN bonds and $=0$ on even ones. The odd bonds take large values:
$\{-2.590$, $-2.183$, $-2.063$, $-2.018$, $-2.018$, $-2.063$,
$-2.183$, $-2.590\}$. This perturbative result uncovers a tendency
of the system towards spin dimerization. The transition to the dimerized
state is discontinuous (first-order) by the very nature of the degenerate
perturbation theory.

However, the degenerate perturbative treatment is not justified when
$\lambda$ becomes comparable or stronger than the orbital gap $\simeq1/L$,
hence it must fail, at least quantitatively, in the thermodynamic
limit. This is why in the following we use DMRG for OBC to obtain
numerically exact results in this limit.

\section{Crossover to the thermodynamic limit: ${\bf \lambda\simeq1/L}$\label{sec:crossover}}
In our DMRG calculation we used the dynamic
block-state selection algorithm. \cite{DBSS:cikk1,DBSS:cikk2} The a priori value of the quantum
information loss was set to $\chi=10^{-7}$, which required to
keep block states up to 4000. Typical truncation errors were in order of $10^{-8}$. We considered
finite chains with OBC up to length $L=800$.
\par The
thermodynamic limit is reached
when the system size is longer than its correlation length $\xi$.
The crossover to this regime can be seen e.g. in the entanglement (von Neumann)
entropy, $s(\lambda,L/2)$, between the two halves of the
chain.\cite{vidallatorre03,legeza2003b,calabrese04} This quantity 
can be obtained from the appropriate reduced density matrix, $\rho_{L/2}(\lambda)$:
\begin{equation}
 s(\lambda,L/2)=-{\rm Tr}\rho_{L/2}(\lambda)\ln\rho_{L/2}(\lambda),
\end{equation}
where $\rho_{L/2}(\lambda)$ denotes the reduced density matrix of the subsystem containing sites
$1,...,L/2$.
In Fig.
\ref{fig:block_entropy_scaling}, we show the entropy for different
system sizes and small $\lambda\ll1$.

For small system sizes, or large $1/L$, the entropies for different
$\lambda$ collapse. This is the regime of validity of the degenerate
perturbation theory, where the actual strength of a weak $\lambda\ll1/L$
is irrelevant.

For each $\lambda$ the entropy is growing with the system size $L$
until it saturates below a crossover value of $1/L$. This saturation
is characteristic for the thermodynamic limit. Interestingly, the
weaker $\lambda$ is the longer $L$ is needed for the saturation,
hence the correlation length $\xi$ must increase or even diverge
when $\lambda\to0$. In the following Section, we consider mainly
`saturated' results in the thermodynamic limit for all physical observables.

\section{Thermodynamic limit: second-order transition for ${\bf \lambda\to0^{+}}$\label{sec:weaklambda}}

\begin{figure}[t!]
\includegraphics[clip,width=1\columnwidth]{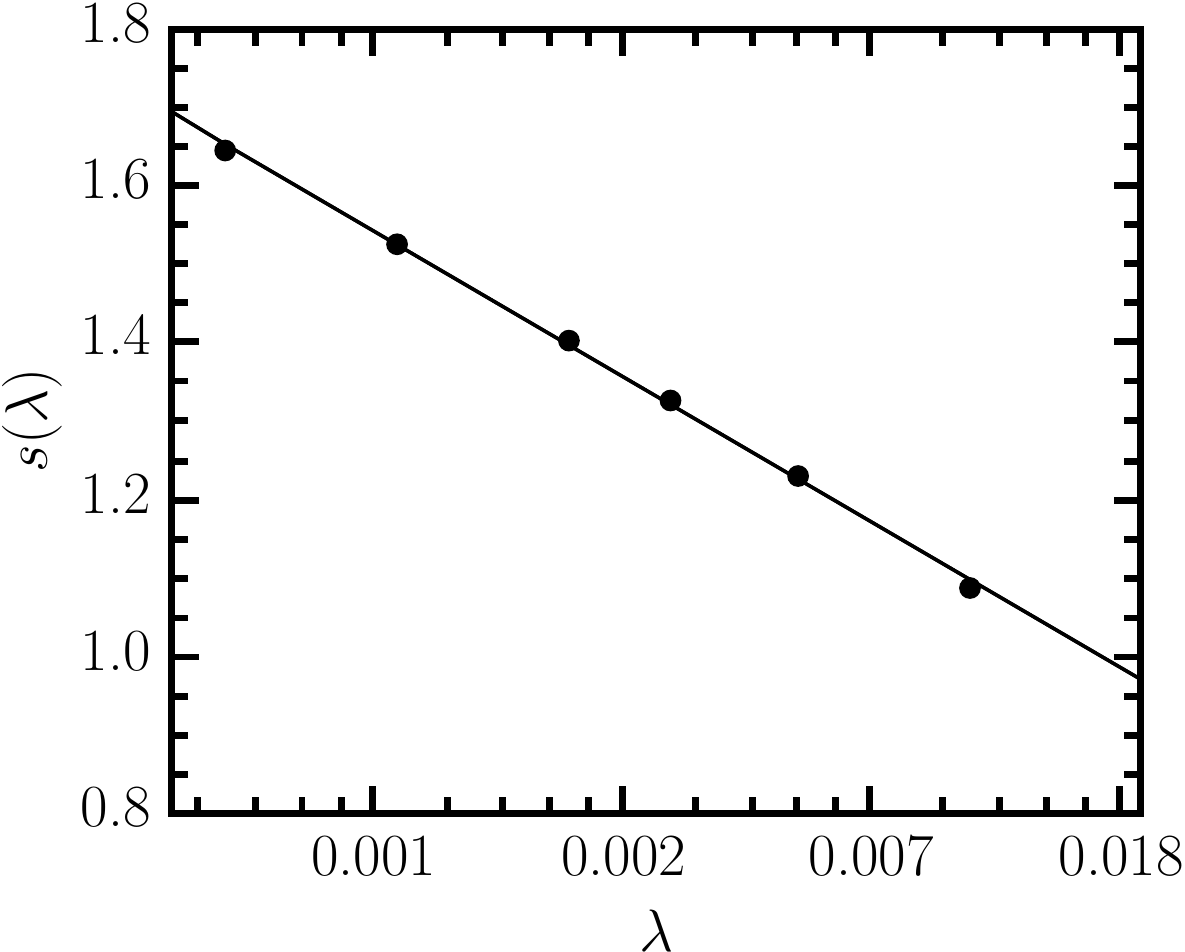}
\caption{\label{fig:block_entropy_lambda} The entanglement entropy of a half-chain
as a function of $\lambda$ in the thermodynamic limit. Note that
the $x$-scale is logarithmic. The points are the DMRG results, while
the solid line is the best fit using Eq. (\ref{eq:entropy_scaling}). }
\end{figure}

Firstly, we consider the behavior of the block entropy of a half-chain, $s(\lambda)$, in the thermodynamic limit: 
\begin{equation}
s(\lambda)=\lim_{L\rightarrow\infty}s(\lambda,L/2).
\end{equation}
This is shown in Fig. \ref{fig:block_entropy_lambda}, where $x$-scale is logarithmic.  

The entropy diverges as we approach the critical point. Near the critical
point at $\lambda\to0^{+}$ the entropy is expected to behave as \cite{calabrese04}
\begin{equation}
s(\lambda)=\frac{c}{6}\ln(\xi)=-\frac{c\nu}{6}\ln\left(\lambda/\lambda_{0}\right),
\label{eq:entropy_scaling}
\end{equation}
where $c$ is the central charge and $\xi=(\lambda/\lambda_{0})^{-\nu}$ the correlation length with the exponent $\nu$. 
The best fit is $\lambda_{0}=3.9\pm0.5$ and $c\nu=1.11\pm0.02$.

\begin{figure}[t!]
\includegraphics[clip,width=1\columnwidth]{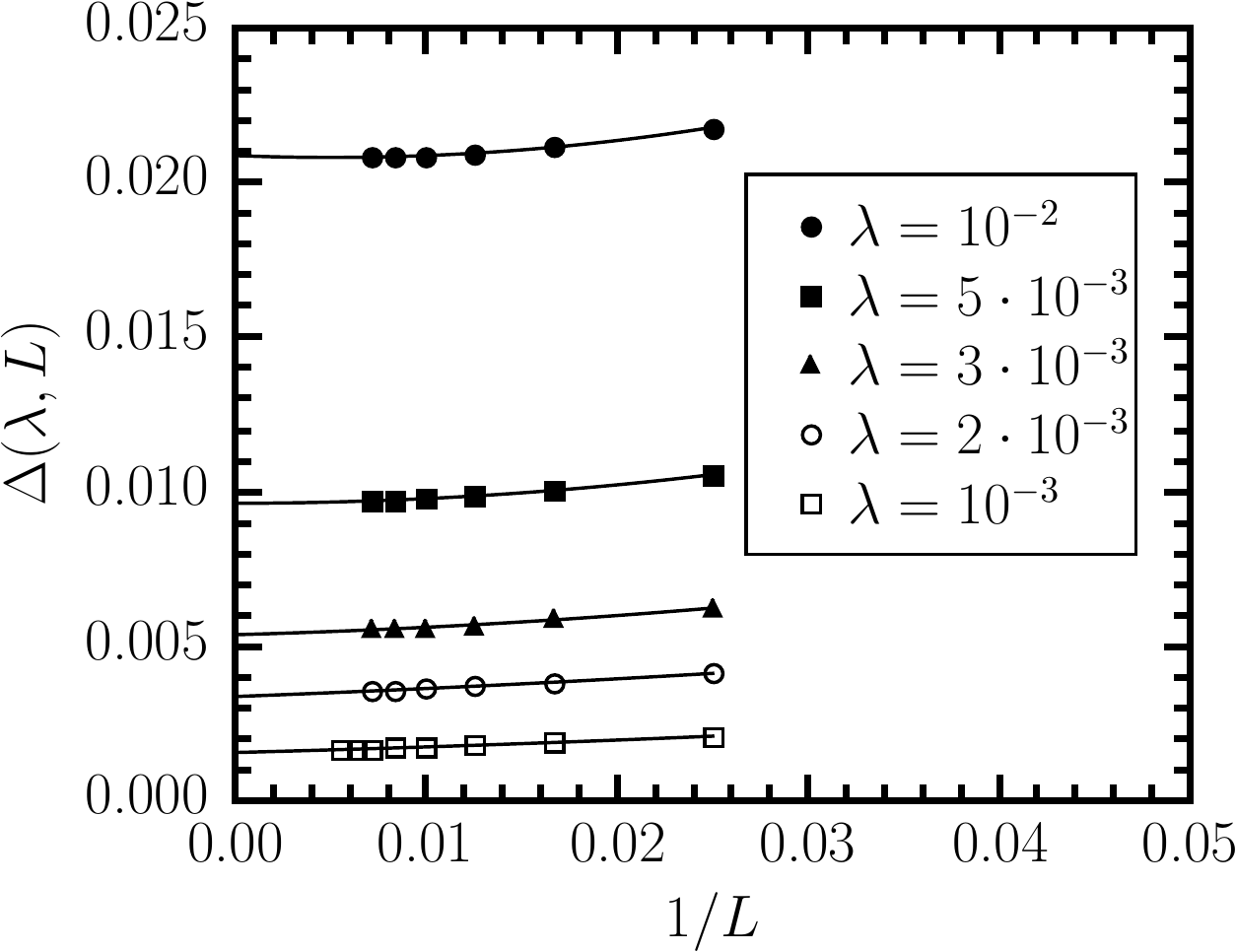}
\caption{\label{fig:gap_scaling} The finite-size scaling of the spin gap for
several values of $\lambda$. The solid lines denote the result of
our fit using Eq. (\ref{eq:gap_ansatz}). }
\end{figure}

The diverging block entropy indicates a diverging correlation length
$\xi$ in the system and a vanishing gap $\Delta=\Delta_{0}\lambda^{z\nu}$,
where $z$ is the dynamical exponent. Therefore we investigated the
first few low-lying excitations in the model. We found that the first
excitation for $\lambda\ll1$ is a spin flip, that is, the smallest
gap is given by the spin gap. To obtain the bulk values of the gap, 
$\Delta(\lambda)$, we performed extrapolation to the thermodynamic 
limit with the fitting ansatz 
\begin{equation}
\Delta(\lambda,L)=\Delta(\lambda)+A/L+B/L^{2},\label{eq:gap_ansatz}
\end{equation}
where $\Delta(\lambda)$, $A$ and $B$ are free parameters. This
ansatz is motivated by the fact that the finite size corrections are
expected to be algebraic for OBC. The finite-size scaling procedure
is demonstrated in Fig. \ref{fig:gap_scaling}. 

\begin{figure}[t!]
\includegraphics[clip,width=1\columnwidth]{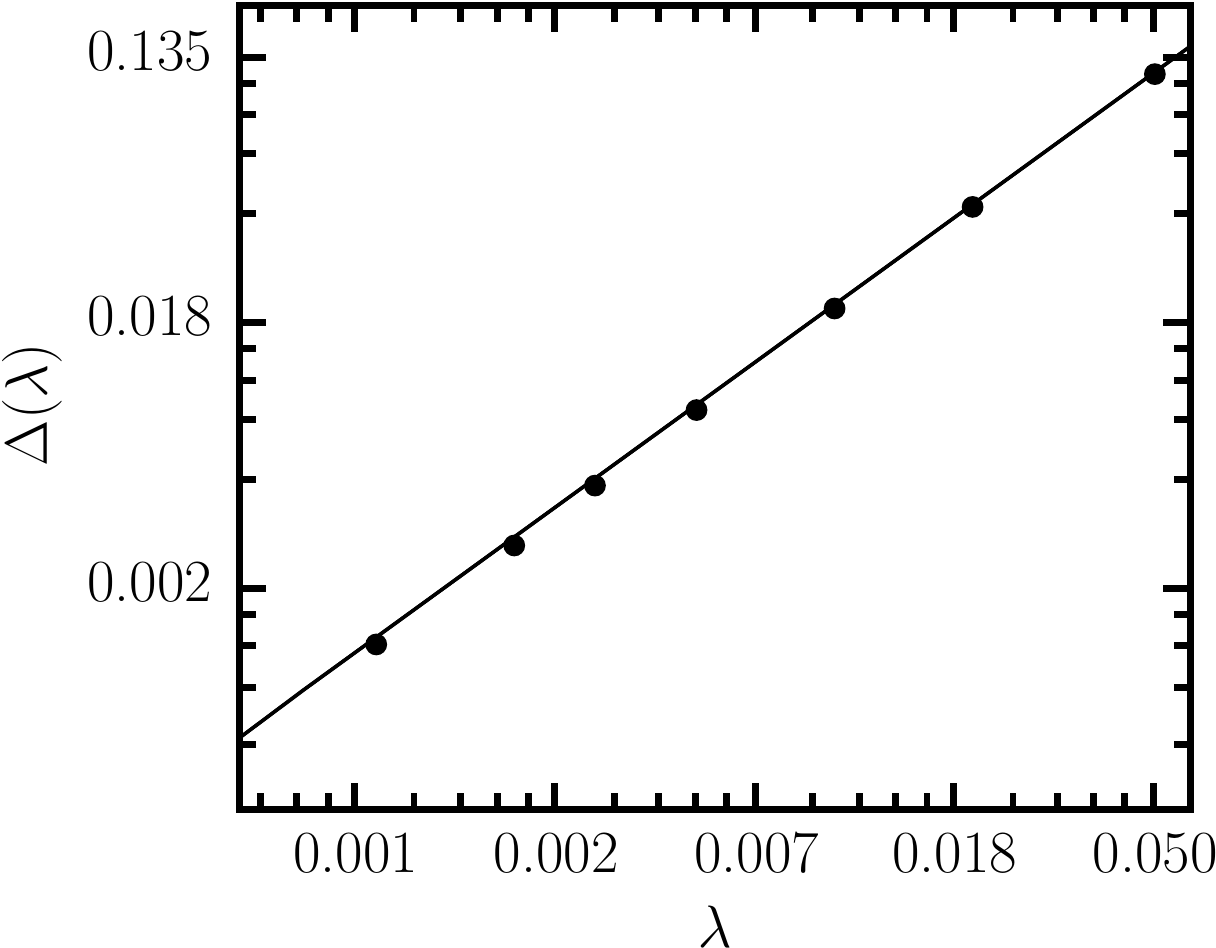} 
\caption{\label{fig:gap_lambda} The spin gap extrapolated to the thermodynamic limit as a function of $\lambda$. Note that both scales are logarithmic.
The points are the DMRG results, while the solid line is the best fit using Eq. (\ref{eq:gap_lambda_dependence}). }
\end{figure}

The extrapolated gap as a function of $\lambda$ is shown in Fig. \ref{fig:gap_lambda} with a log-log scale. 

The data points can be fitted well with a linear, that is, the gap exhibits a power-law dependence on $\lambda$: 
\begin{equation}
\Delta(\lambda)=\Delta_{0}\lambda^{z\nu},\label{eq:gap_lambda_dependence}
\end{equation}
where $\Delta_{0}=3.12\pm0.02$ and $z\nu=1.088\pm0.002$. This means
that the gap vanishes continuously in contrast to the naive prediction
of the perturbation calculation, and a second-order phase transition
occurs at $\lambda=0^{+}$.

\begin{figure}[t!]
\includegraphics[clip,width=1\columnwidth]{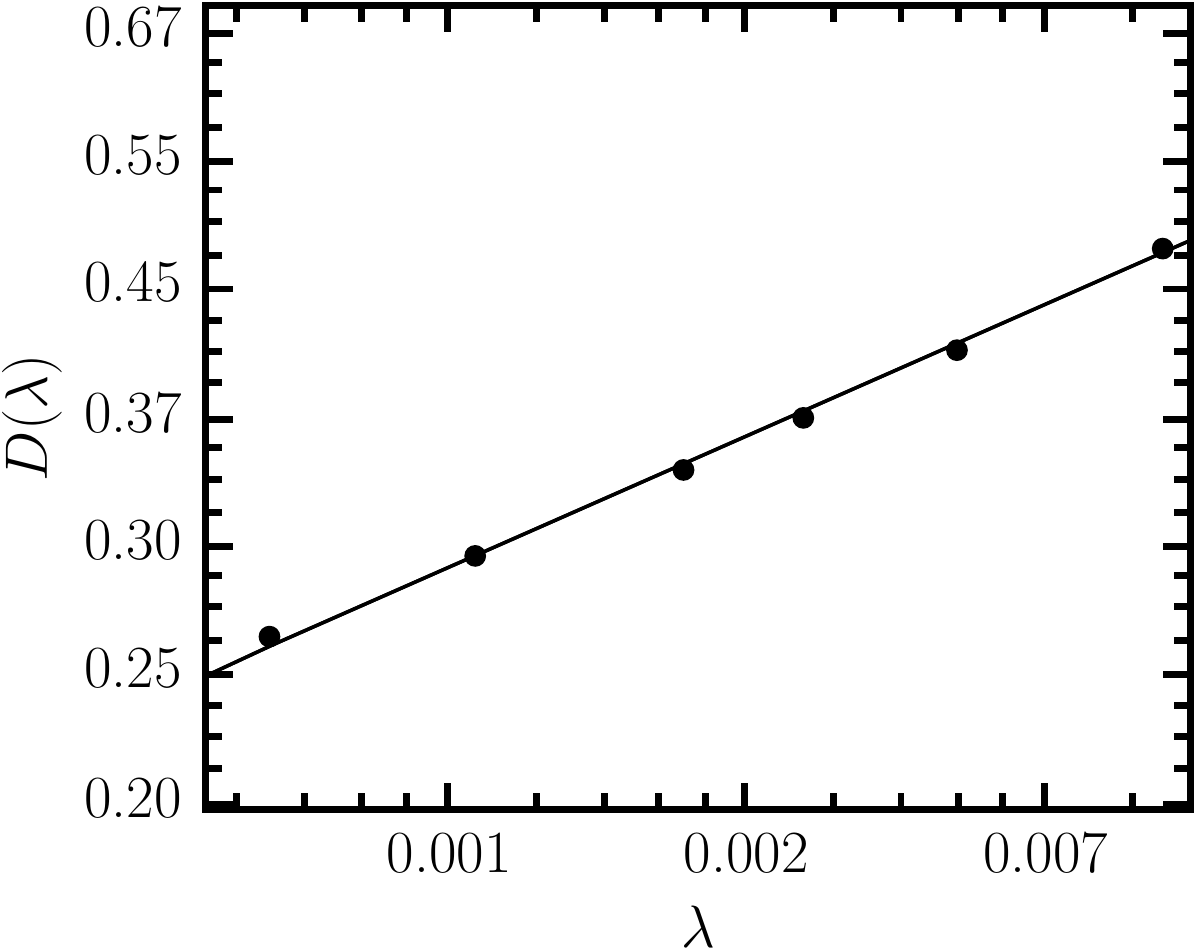}
\caption{\label{fig:dimerization_lambda} The extrapolated dimerization order
parameter, $D$, as a function of $\lambda$. The points are the DMRG
results, while the solid line denote the fit using Eq.
(\ref{eq:dimerization_lambda_dependence}). Note that both scales are logarithmic.}
\end{figure}

The same applies for the dimerization, $D$, which is defined as 
\begin{equation}
D(\lambda)=\lim_{L\rightarrow\infty}|s(\lambda,L/2)-s(\lambda,L/2-1)|.
\end{equation}
This is shown in Fig. \ref{fig:dimerization_lambda} with a log-log
scale. 
The dimerization -- similarly to the gap -- behaves as 
\begin{equation}
D(\lambda)=D_{0}\lambda^{\beta},\label{eq:dimerization_lambda_dependence}
\end{equation}
where $D_{0}=1.22\pm0.03$ and $\beta=0.204\pm0.004$.

\section{ Dimerization for finite $\lambda$ both large and small \label{sec:Dimerization}}

The previous Section focused on the critical behaviour in the limit
$\lambda\to0^{+}$. This is a second-order phase transition, where
the symmetry broken by dimerization is restored. In this Section we
paint a general picture of the dimerized phase at $\lambda>0$. We
distinguish two regimes, of large and small $\lambda$, where the
nature of the dimerization is qualitatively different. They are separated
by a smooth crossover.
\par
\begin{figure}[t!]
\includegraphics[clip,width=1\columnwidth]{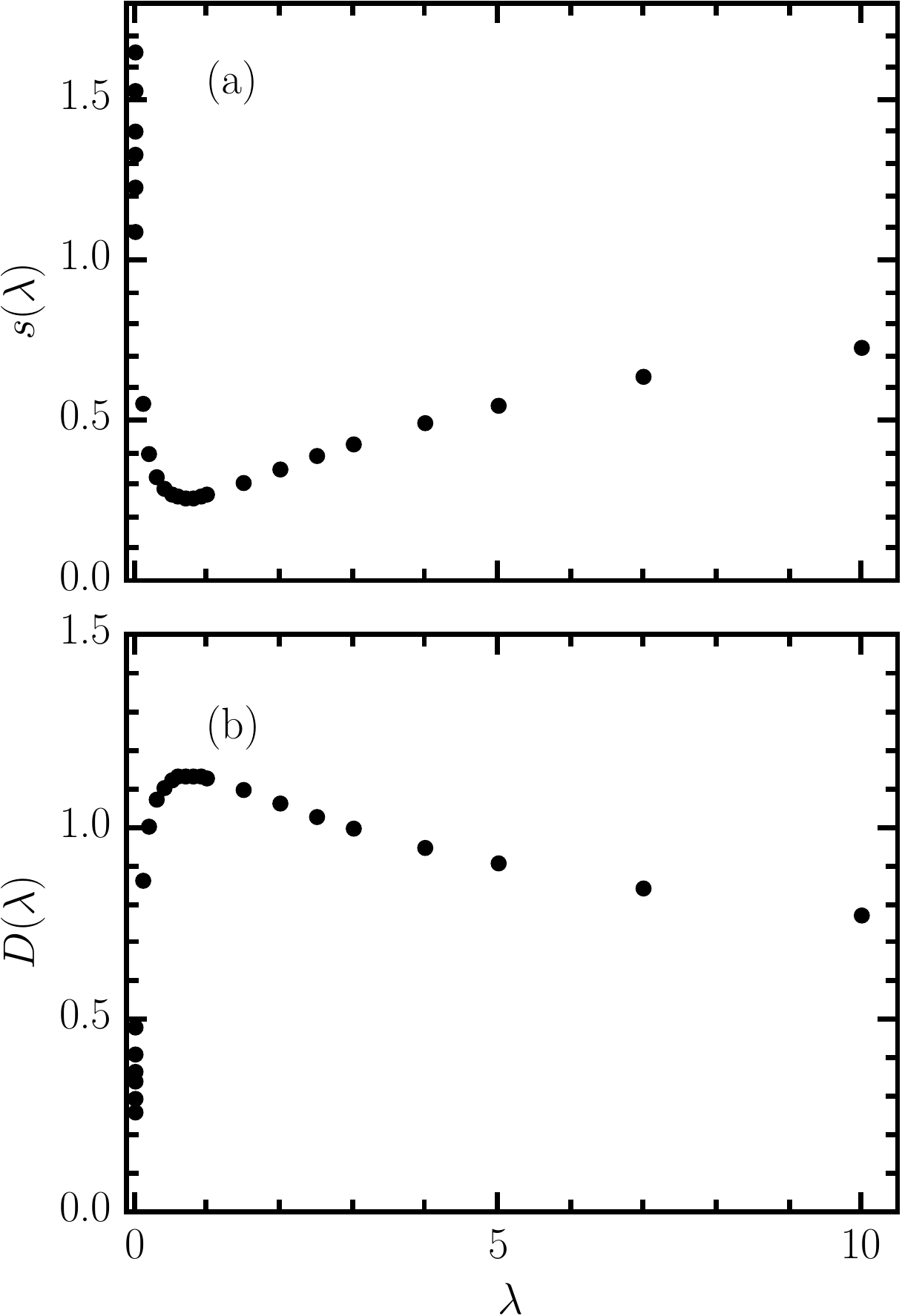}
\caption{\label{fig:dim_entropy_fig} The extrapolated block entropy, $s(\lambda)$, and
dimerization order
parameter, $D(\lambda)$, as a function of $\lambda$ are shown in panels (a) and (b), respectively. }
\end{figure}

 It is interesting to analyze the behavior of the block entropy and the dimerization order
parameter for larger $\lambda$ before going into the details of the dimerized phase. They are shown
in Fig. \ref{fig:dim_entropy_fig}. 
It is seen that around $\lambda\simeq0.7$ the block entropy and the dimerization order parameter
exhibits a local minimum and maximum, respectively. This can be understood as follows: when the
dimerization is the strongest the ground state is the less entangled, which manifests in the minimum
value of the block entropy. We address what happens for $\lambda\to\infty$ later in this
section.

\subsection{Numerical findings\label{sub:num}}

\begin{figure}[t!]
\includegraphics[clip,width=1\columnwidth]{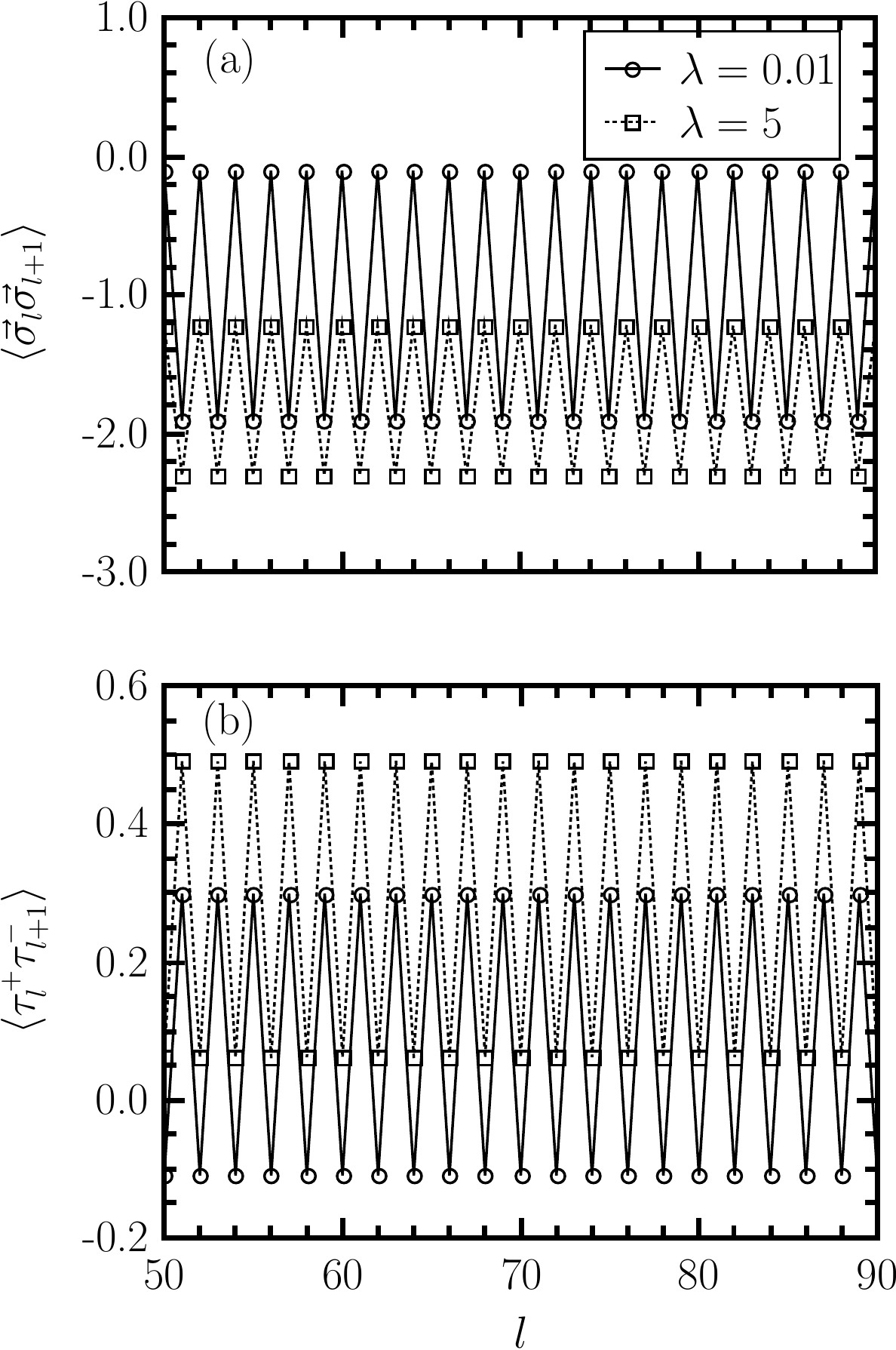}
\caption{\label{fig:NNcorr_obc} DMRG results for the NN spin (panel (a)) and
orbital (panel (b)) correlations for the open chain of $L=140$ sites
as a function of the site number $l$ for $\lambda=0.01$ and $\lambda=5$.}
\end{figure}

Numerical observations are as follows. For any $\lambda$ we observe
both spin and orbital dimerization in the sense that the NN correlations
alter between large/small values for odd/even bonds as shown in Fig.
\ref{fig:NNcorr_obc}. The difference between spin and orbital sector
is that the spin correlations remain always AF while the orbital ones
can alternate between positive/negative values for small $\lambda$
and remain FM only for large $\lambda$.
The behavior of the long-range spin and orbital correlations is different
in the small and large $\lambda$ regime. While the correlation $\langle\vec{\sigma}_{l}\vec{\sigma}_{l+R}\rangle$
always decays exponentially with $R$ it remains constantly
negative for small $\lambda$ while for large $\lambda$ it alternates
between negative/positive values. Thus we say that for small $\lambda$
we have strong dimerization of the spins. Because of the
negative sign of the correlations this is a spin-singlet dimerization.
The behavior of the orbital correlations seems to be complementary,
the long range correlations $\langle\tau_{l}^{+}\tau_{l+R}^{-}\rangle$
are constantly positive for large $\lambda$, while for small
$\lambda$ they alternate with $R$. In both regimes they decay exponentially
with $R$. Thus, analogically to the spin sector, we say that for
large $\lambda$ the orbitals exhibit strong dimerization
in the orbital-triplet state. Summarizing, we have found strong spin-singlet
dimerization for small $\lambda$ and strong orbital-triplet dimerization
for large $\lambda$.
\begin{figure}[t!]
\includegraphics[clip,width=1\columnwidth]{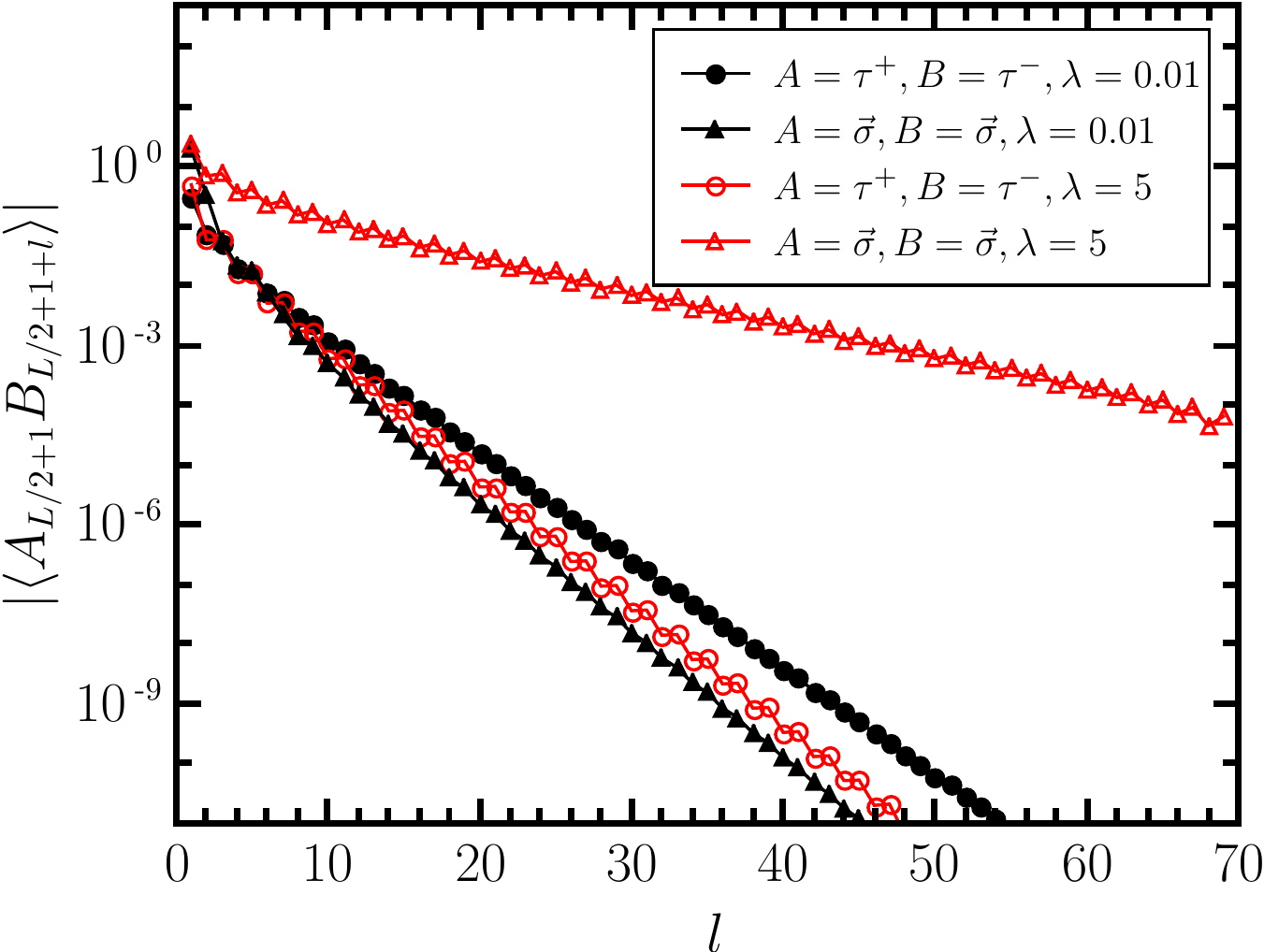} 
\caption{\label{fig:LRcorr_obc} DMRG results for the spin and orbital correlations
for the open chain of $L=140$ sites as a function of the site number
$l$ for $\lambda=0.01$ and $\lambda=5$. The vertical axis has a
logarithmic scale. }
\end{figure}
\begin{figure}[t!]
\includegraphics[clip,width=1\columnwidth]{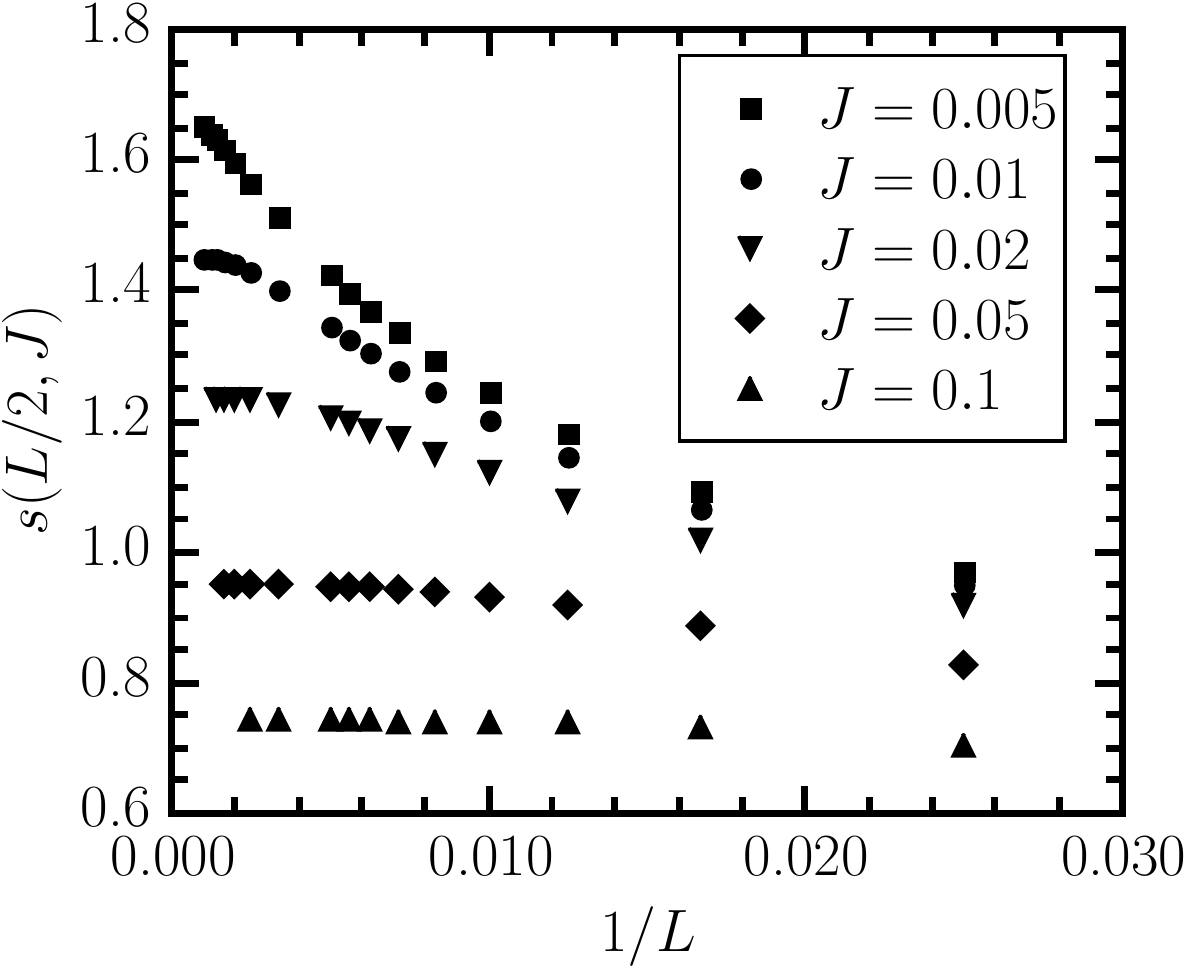}
\caption{\label{fig:block_entropy_scaling2} Finite-size scaling results for
the block entropy of an open chain as a function of $1/L$  (up to $L=800$) for several
decreasing values of $J$ and $\lambda=1$. For each $J$ the entropy saturates for
small enough $1/L$ where the thermodynamic limit is achieved.}
\end{figure}

Finally, we note that the transition between the two regimes, of strong
dimerization of either spins or orbitals, is a crossover without any
features of a phase transition. However, one can easily observe a
change in the behavior of the entanglement entropy of the half-system
$s(\lambda)$ in Fig. \ref{fig:dim_entropy_fig} (a); while it is rapidly decreasing for small
$\lambda$, it becomes
slowly increasing for larger values of $\lambda$ with a shallow minimum
at $\lambda\simeq0.7$, where the dimerization is the strongest.

At the same time, in addition to the qualitative change of the spin
correlation triggered by increasing $\lambda$, there is a big difference
in their correlation lengths. They are shown in Table \ref{table:I}.
From the semilogarithmic plot of Fig. \ref{fig:LRcorr_obc} we can learn that
for small $\lambda$ both correlations decay with almost the same
exponent, while for large $\lambda$ the spin correlation function
has much larger correlation length than that of the orbital correlation
function. This suggests that there is a second phase transition
in the system located at $\lambda\to\infty$ or $J=0$. This is not
totally unexpected as in this limit the KH model becomes
simply the Heisenberg model in the spin sector which is known to be critical.
This observation is confirmed by the behavior of the block entropy
of the half system, $s(L/2,J)$, which exhibits  similar scaling properties
as $J\to0$ for $\lambda=1$, see Fig.
\ref{fig:block_entropy_scaling2}, to what we have seen for $\lambda\to0$ in Eq.
(\ref{eq:entropy_scaling}). In the present case we fit the product
$c\nu=1.85\pm0.01$. Since $c=1$ for the Heisenberg chain, 
we obtain an estimate for the correlation length exponent $\nu=1.85\pm0.01$.

\begin{table}[h]
\begin{tabular}{@{}c@{\hspace{4mm}}c@{\hspace{4mm}}c@{\hspace{4mm}}}
\toprule 
$\lambda$  & $\xi_{{\rm orbital}}$  & $\xi_{{\rm spin}}$\tabularnewline
\midrule 
$0.01$  & $5.52$  & $4.65$\tabularnewline
$5$  & $4.72$  & $18.18$ \tabularnewline
\bottomrule
\end{tabular}
\caption{The correlation lengths of the spin and orbital correlation functions
for several values of $\lambda$.}
\label{table:I} 
\end{table}

\subsection{Qualitative explanation\label{sub:anal}}

The dimerization of spins involving constantly negative long-range
correlation function can be understood in terms of the first order
perturbative expansion of Sec. \ref{sec:pert} and especially looking at 
Fig. \ref{fig:ladder}. As we can see, in the Kumar basis the singlet 
bonds are mainly on the ladder's rungs connecting spins at sites $N+i$ 
and $N-i+1$. In the physical basis the rung singlets become, in first approximation, 
spin singlets on every second bond. In this approximation the correlation function 
is strongly negative only within a given singlet and zero anywhere else. 
However,
when we take into account fluctuating positions of the up- and down-orbitals in the orbital Fermi sea,
then the singlets on every second bond become smeared over neighboring sites.
This smearing implies that: 
($i$) singlets placed every second bond are not perfect, i.e. $0>\langle \vec{\sigma}_{l}\vec{\sigma}_{l+1} \rangle>-3$  
and ($ii$) there are negative spin correlations between sites $l$ and $m\in[l-\delta l,l+\delta l]$ where $\delta l$ is the 
legth scale at which a single up orbital is delocalized within the
Fermi sea. In this framework the strongly spin-dimerized ground
state of the KH model can be understood as a spin liquid of strongly 
resonating singlets.  

The triplet dimerization of the orbitals for large $\lambda$ can
be easily understood in the mean-field way similarly as it was done
in Ref. \cite{PhysRevLett.101.157204}. In this limit the leading
part of the Hamiltonian is a pure Heisenberg term ${\cal V}$ for
the spins. Thus we can use a mean-field decoupling of a spin-orbital
term in ${\cal H}_{0}$: 
\begin{eqnarray}
X_{l,l+1}\!\left(\tau_{l}^{+}\tau_{l+1}^{-}\!+\!\tau_{l}^{-}\tau_{l+1}^{+}\right)\approx\left\langle X_{l,l+1}\right\rangle \left(\tau_{l}^{+}\tau_{l+1}^{-}\!+\!\tau_{l}^{-}\tau_{l+1}^{+}\right)+\nonumber \\
X_{l,l+1}\left\langle \tau_{l}^{+}\tau_{l+1}^{-}\!+\!\tau_{l}^{-}\tau_{l+1}^{+}\right\rangle -\left\langle X_{l,l+1}\right\rangle \left\langle \tau_{l}^{+}\tau_{l+1}^{-}\!+\!\tau_{l}^{-}\tau_{l+1}^{+}\right\rangle .\nonumber \\
\label{eq:so_deco}
\end{eqnarray}
The spins are AF with $\langle\vec{\sigma}_{l}\vec{\sigma}_{l+1}\rangle<-1$,
hence for each bond we have $\langle X_{l,l+1}\rangle<0$ and, according
to the first line of Eq. (\ref{eq:so_deco}), the orbitals order ferromagnetically.
The instability towards dimerization, meaning here formation of the
orbital triplets, can be demonstrated by imagining a self-consistecy
loop for a non-uniform MF Hamiltonian. If, for any reason, a bond
has increased the orbital correlation, then, according to the second
line of Eq. (\ref{eq:so_deco}), the spins will feel an increased
tendency towards antiferromagnetism. Then again, the first line of
Eq. (\ref{eq:so_deco}) implies that the orbital bonds will be increased
and so the spin bond will follow. Such a loop clearly leads to maximalization
of the orbital correlations at some bonds and, since we cannot have
a triplet at every bond, we must have the dimerization of the orbitals.
We verified this argument by solving numerically a non-uniform MF
Hamiltonian for $L=16$.

\section{Conclusion\label{sec:Summa}}

\begin{figure}[t!]
\includegraphics[clip,width=1\columnwidth]{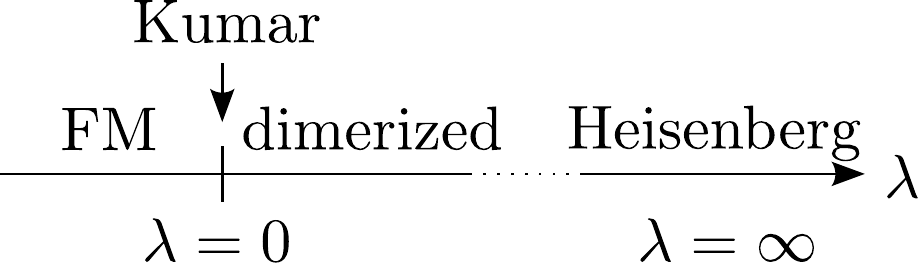}
\caption{\label{fig:phase_diag} Schematic phase diagram of the Kumar-Heisenberg model for $J=1$.}
\end{figure}

We have found that an admixture of the Heisenberg interactions $\lambda$
between the spins causes a spontaneous dimerization of the Kumar model. 
The dimerization is both in the spin and orbital sector but it 
becomes strong for the spins in the regime of small $\lambda$ and for
the orbitals in the opposite limit. We have shown that the dimerization
of spins can be understood by means of the Kumar transformation, i.e., 
in a basis where the spins and orbitals in the Kumar Hamiltonian get decoupled.
On the other hand we show that the dimerization of orbitals for large $\lambda$
can be understood be a mean field mechanism in the physical basis.    

We have seen that the perturbative expansion in the Kumar basis, however 
useful for understanding the spin dimerization, fails to capture the 
second-order phase transition found by DMRG in the thermodynamic limit.
Whereas the perturbative prediction is always a discontinuous transition, 
the numerical study of the system of sizes up to $L=600$ allowed us to 
observe the divergence of the block entropy, vanishing of the gap and 
the dimerization order parameter in the regime where we approach 
the transition point at $\lambda=0$ {\it slower} than the gap in the
pure Kumar model closes, i.e., in the regime where the perturbative
approach must fail. Finally, by observing that the spin correlation
length is strongly increased as $\lambda$ grows we have deduced that
there is another phase transition in the limit of $\lambda\to\infty$
(or equivalently $J=0$). This we have confirmed by the finite-size scaling 
of the block entropy for decreasing $J$. The phase diagram is shown schematically in Fig.
\ref{fig:phase_diag}.

Finally we note that however the full dimerization in orbital space
can be easily understood by the mean-field approach, similarly as
it was done for YVO$_{3}$ \cite{PhysRevLett.101.157204} at finite
temperature, but here in zero temperature, the spin dimerization is
more complex and its mechanism can be revealed only by the Kumar transformation.
This is a novel feature that was not found in the similar spin-orbital
models, namely SU(2)$\otimes$XY \cite{PhysRevB.61.5868} and SU(2)$\otimes$SU(2)
\cite{PhysRevB.61.5868,PhysRevLett.81.5406} one, where in principle
the dimerization could be captured by a simple variational wave functions
in the physical basis and confirmed via DMRG. We argue that such a
novel dimerization can be observed in the (quasi) one-dimensional
compounds with active spin and orbital degrees of freedom.

\acknowledgments  We thank Andrzej M. Ole\'s for insightful discussions.
We acknowledge financial support by the Polish National
Science Center (NCN) under Projects No. 2012/04/A/ST3/00331 (W.B.)
and 2013/09/B/ST3/01603 (J.D.). W.B. was also supported by the Foundation
for Polish Science (FNP) within the START program. Ö.L. and I.H. were
supported in part by the Hungarian Research Fund (OTKA) through Grant
Nos.~K 100908 (Ö.L. and I.H.) and NN110360 (Ö.L.). The research of
I. H. was supported by the European Union and the State of Hungary,
co-financed by the European Social Fund in the framework of TÁMOP-4.2.4.A/
2-11/1-2012-0001 'National Excellence Program'.

\appendix

\section{The effective spin couplings\label{app:J}}

\begin{figure}[t!]
\includegraphics[width=0.75\columnwidth]{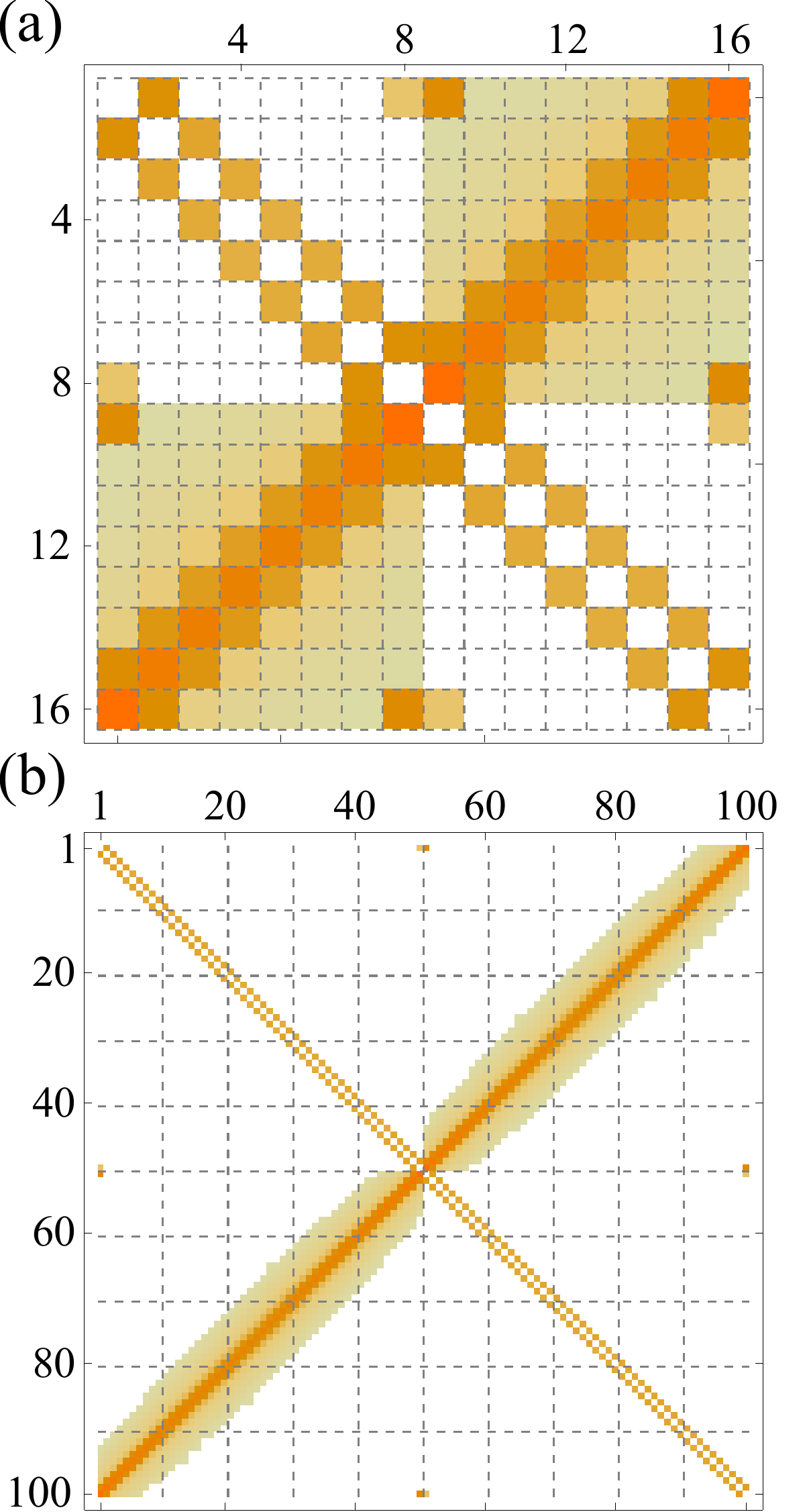}
\caption{Plots of the $\mathbf{J}$ coupling matrices of the effective spin
Hamiltonians $\tilde{{\cal H}}_{1}$ for system sizes; (a)--- $L=16$
and (b)--- $L=100$. Colors tending to red mean high positive values,
white fields mean zero values. Dashed grid is a guide for the eye.
\label{fig:Jplots}}
\end{figure}

The effective, first-order couplings between the spins can be expressed as, 
\begin{eqnarray}
\mbox{J}_{i,j}=\sum_{p_{1},...,p_{N}}P(\vec{p})\sum_{l=1}^{N}\delta_{i,f(l,\vec{p})}~\delta_{j,f(l+1,\vec{p})},
\end{eqnarray}
where the first sum is an average over a probability distribution
\begin{equation}
P(\vec{p})=\prod_{1\leq i<j\leq N}\left\{ 4\sin^{2}\frac{\pi\left(p_{j}-p_{i}\right)}{L}\right\} ,\label{eq:det}
\end{equation}
for empty sites $\{p_1,...,p_N\}$ in a half-filled Fermi sea. 
Here we consider periodic boundary conditions to minimize boundary effects.
For $N=L/2$ the permutation maps the $N$ consecutive spins on empty sites (orbitals up) to spins
$N,...,1$ and the $N$ consecutive spins on occupied sites (orbitals down) to spins $N+1,...,2N$. 
In the physical representation only nearest-neighbor spins are coupled. 
After the transformation ${\cal U}$ each physical NN coupling (PNNC) contributes to (is smeared over) many $\mbox{J}_{i,j}$.

The matrix $\mbox{J}_{i,j}$ is shown in Figs. \ref{fig:Jplots}(a) and \ref{fig:Jplots}(b) as color plots for a small and large $L$ respectively. 
The matrix has two diagonal and two off-diagonal $N\times N$ blocks. 
The diagonal block $i,j=1,...,N$ originates from PNNC's between spins on empty sites. 
Since only a pair of consecutive empty sites can happen to be NN's, the block couples only NN's. 
The weakest coupling $\mbox{J}_{1,N}$ originates from the relatively rare situation when
the first and last empty sites happen to be also respectively the first and last sites of the chain, i.e., 
they are not only NN's (across the closing bond) but also each of them occupies a definite position.
$\mbox{J}_{1,N}=0$ for an open chain. 
The diagonal block $i,j=N+1,...,2N$ originates from PNNC's between spins on occupied sites. 
In our half-filled orbital Fermi sea the two diagonal blocks are the same.

The two identical off-diagonal blocks originate from PNNC's between pairs of empty and occupied sites. 
The top values of the coupling are between the first and the last site: 
$\mbox{J}_{1,L}\simeq0.705$ for $L=16$ and $\mbox{J}_{1,L}\simeq0.783$ for $L=100$. 
This strong coupling originates from the PNNC between the $N$-th empty and the $N$-th occupied site that are very likely to be next to each other. 
The coupling $\mbox{J}_{1,N+1}$ originates from the PNNC between the $N$-th empty and the $1$-st occupied site that is possible mainly across the closing bond. 
This coupling is much weaker for an open chain where the $N$-th empty and $1$-st occupied sites are rather unlikely to be NN. 
However, 
the dominant feature of each off-diagonal block is its antidiagonal belt. 
A coupling $\mbox{J}_{N-i,j}$ within this belt originates from a PNNC between $i$-th empty and $j$-th occupied site with $i$ close to $j$. 
These sites are most likely to be NN when their difference $|i-j|$ is not much greater than a variation of a position of the $i$-th empty 
(or the $j$-th occupied) site in the Fermi sea. 
This variation limits the width of the antidiagonal belt.

We found that truncating the full matrix $\mbox{J}_{i,j}$ to its dominant antidiagonal belt is a very good approximation. 
We are assuming this approximation in the main text.

\end{document}